\documentclass[conference]{IEEEtran}
\IEEEoverridecommandlockouts
\usepackage{cite}
\usepackage{amsmath,amssymb,amsfonts}
\usepackage{algorithmic}
\usepackage{lipsum}
\usepackage{graphicx}
\usepackage{amsmath}
\usepackage{bbm}
\usepackage{booktabs}
\usepackage{multirow}
\usepackage[normalem]{ulem}
\ifCLASSOPTIONcompsoc
    \usepackage[caption=false, font=normalsize, labelfont=sf, textfont=sf]{subfig}
\else
\usepackage[caption=false, font=footnotesize]{subfig}
\fi
\usepackage{textcomp}
\usepackage{xcolor}
\def\BibTeX{{\rm B\kern-.05em{\sc i\kern-.025em b}\kern-.08em
    T\kern-.1667em\lower.7ex\hbox{E}\kern-.125emX}}

\usepackage{comment}

\begin{document}

\title{Age of Loop for Wireless Networked Control Systems Optimization
}

\author{\IEEEauthorblockN{Pedro M. de Sant Ana\IEEEauthorrefmark{1},
Nikolaj Marchenko\IEEEauthorrefmark{1}, 
Petar Popovski\IEEEauthorrefmark{2} and
Beatriz Soret\IEEEauthorrefmark{2}}
\IEEEauthorblockA{\IEEEauthorrefmark{1}Corporate Research and Advanced Engineering, Robert Bosch GmbH, Renningen, Germany\\
Email: \{Pedro.MaiadeSantAna, Nikolaj.Marchenko\}@de.bosch.com
}
\IEEEauthorblockA{\IEEEauthorrefmark{2}Department of Electronic Systems, Aalborg University, Aalborg, Denmark\\
Email: \{petarp, bsa\}@es.aau.dk
}
}
\maketitle

\begin{abstract}
Joint design of control and communication in Wireless Networked Control Systems (WNCS) is a promising approach for future wireless industrial applications. In this context, Age of Information (AoI) has been increasingly utilized as a metric that is more representative than latency in the context of systems with a sense-compute-actuate cycle. Nevertheless, AoI is commonly defined for a single communication direction, Downlink or Uplink, which does not capture the closed-loop dynamics.
In this paper, we extend the concept of AoI by defining a new metric, \emph{Age of Loop (AoL)}, relevant for WNCS closed-loop systems. The AoL is defined as the time elapsed since the piece of information causing the latest action or state (depending on the selected time origin) was generated. We then use the proposed metric to learn the WNCS latency and freshness bounds and we apply such learning methodology to minimize the long term WNCS cost with the least amount of bandwidth. We show that, using the AoL, we can learn the control system requirement and use this information to optimize network resources. 
\end{abstract}

\IEEEpeerreviewmaketitle

\section{Introduction}
\label{intro}
%Wireless Networked Control Systems (WNCS) are seen as an essential enabler for future industrial, logistics, and transport applications, where they can provide a high level of flexibility, data fusion, resource sharing, and cost reduction. 

Networked control systems (NCS) are an essential part of many industrial domains such as  factory automation, logistics, or transportation. Wireless NCSs (WNCS) enable mobile control applications were wiring is not possible or high flexibility is required. However, due to the nature of the wireless medium, reliability of WNCS remains an open challenge, in particular for low-latency applications. 

Decisions taken at a system level can have a direct impact on the communication medium and vice-versa, formulated by Witsenhausen as counterexample for distributed control problem \cite{witsenhausen1968counterexample}. In recent work on WNCS, authors have been increasingly exploring the inter-relation between the control and communication schemes. In \cite{liu2020latency} and \cite{gatsis2020latency}, for example, the authors demonstrate how the latency and reliability trade-off directly impacts the system level stability, proposing a co-design of both control and communication entities. Specifically in \cite{liu2020latency}, authors have demonstrated a counter-intuitive conclusion that the plant can still be stabilized with an arbitrarily large delay under certain channel conditions. Another interesting finding was presented in \cite{huang2020wireless}, where authors elucidate an example how one can optimize long-term system performance by assuming more risks with less reliable transmissions in exchange for lower latency. The authors of this publication analyzed the remote AGV control problem in \cite{de2020wireless}.

%\NM{not sure about the paragraph below.} \pp{PP: I agree with Nikolaj. Telling that that latency from 3GPP and 5G-ACIA is inefficient, is not correct, as both organizations now look at otehr timing measures as well. Beatriz has recently looked in the literature for this and can provide you with an input. The best would be to say that the traditional notion of latency of the radio link, in which we attempt to allocate a small, almost deterministic, part of the latency budget to the radio link may not be the most efficient way to go from a system viewpoint. That is why other timing measures have been defined, also in 3GPP and 5G-ACIA, such as age.}

%These findings represent a the fundamental idea that establishing strict latency and reliability requirements for WNCS applications, as discussed in 3GPP in \cite{3gpp_rel_15} and 5G-ACIA in \cite{5gacai}, might be inefficient for WNCS. Evidently, extreme low latency under high reliability rates are enough to guarantee high probability of system level stability, but the main question is if such strict requirements are indeed a necessity, especially considering a huge variety of control systems. 

These findings outline that the traditional notion of latency of the radio link, in which we attempt to allocate a small, almost deterministic, part of the latency budget to the radio link may not be the most efficient way to go from a system viewpoint. That is why other timing measures have been defined, also in 3GPP \cite{3gpp_rel_15} and 5G-ACIA \cite{5gacai}, such as the survival time and recovery time.

In other words, the conventional system design for low-latency and high reliability leads to over-provisioning communication network resources by  decoupling the control and communication entities. Alternatively, the Age of Information (AoI)~\cite{bedewy2019minimizing} has been proven to be a more representative metric for systems with a sense-compute-actuate cycle like the ones considered in this paper, where the receiver is interested in fresh knowledge of the remotely controlled system, rather than the individual packet delay. The AoI defines the time that has elapsed since the newest update available at the destination was generated at the source, and it captures not only the communication delays but also the impact of the packet generation at the controlled process. The other observation regarding the use of resources is that, from the perspective of a Base Station (BS) situated at the controller, WNCS have been traditionally optimized separately for the two transmission directions, uplink (UL) and downlink (DL). 
%\pp{PP: Define exactly what UL and DL is here; you can just say that UL and DL are defined considering that the BS is at the controller.} 
However, WNCS applications and many others are inherently two-way, such that there is a closed-loop where the UL communication can affect the DL and vice-versa, impacting either on system performance or in the use of network resources. In this context, the goal of our paper is to learn the optimal configuration of the communication network that ensures stability of the control system. For this purpose, we consider the two-way nature of the problem and the control-communications interplay. 

%\pp{PP: Can you put a statement here that makes it plausible to see that optimizing for each of the two directions separately is not as efficient as optimizing for the loop?}

%\textit{Could the network automatically learn the control system behavior and use the accomplished information to optimize its resources}?}
This paper contains two main contributions: 
\begin{enumerate}
    \item We propose a new metric, the Age of Loop (AoL), which extends the current AoI definition to take into consideration both UL and DL of the control loop in WNCS, and thus can provide a more precise system state estimation.
    %\NM{Repetition: \sout{Previous work, as \cite{huang2020wireless, klugel2019aoi, champati2019performance}, studied AoI in WNCS applications, but they define the simplest case of a one-way communication link, such that we have a certain AoI measure for the DL communication and another independent AoI measure for the UL communication.} %However, WNCS systems like the one considered in this paper are inherently two way with a closed-loop where the UL communication can affect the DL and vice-versa, impacting either on system performance or in the use of network resources. 
%Such conventional one-way AoI measure is not meaningful in the WNCS closed-loop dynamics.
\item We demonstrate how to apply the AoL metric for joint WNCS optimizations with the application example of a remotely controlled inverted pendulum system~\cite{barto1983neuronlike}.
With a Reinforcement Learning (RL) approach, we find the bandwidth allocation policy based on the AoL state,  which significantly outperforms policies based on fixed latency requirements.
\end{enumerate}
%COMMENT: More details and something about RL here... 
%Then, we consider the novel AoL definition to propose a learning approach to quantify the system performance based on the AoL state. Thereafter we elaborate on the proposed learning methodology to create a bandwidth allocation policy for a remote controlled inverted pendulum system, showing that we can use the learned system requirement to optimize resource allocation.}

The rest of this paper is organized as following: in the next two sections, we introduce the system and WNCS model, respectively. In section \ref{aol_definition}, we define the AoL and show how we can evaluate the control system performance using the proposed metric. Finally, in section \ref{prob_form}, we formulate the bandwidth allocation problem and propose a solution, where the results are analyzed in section \ref{res}.%
\section{System Model}
\label{system_model}
%\NM{NM: I do not like the title system model, here since it is clearly just a subsystem of the whole system. Maybe Controlled System Model?}

\begin{figure}[htpb]
    \centering
    \includegraphics[width=1.0\linewidth]{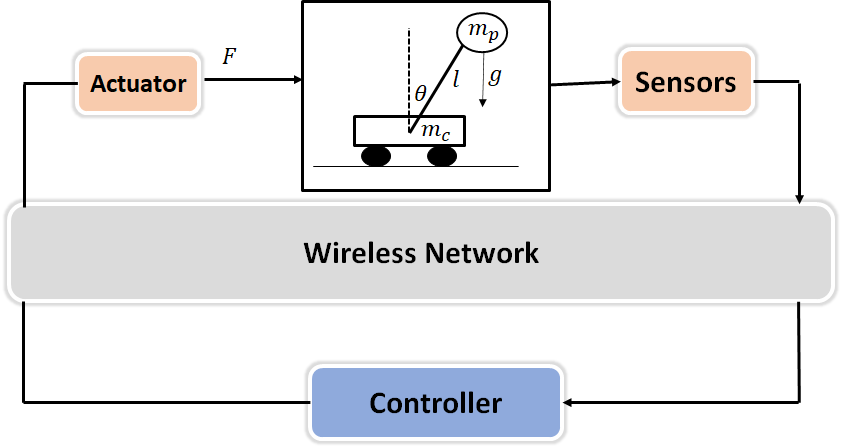}
    \caption{Inverted pendulum system model.}
    \label{fig:cartpole}
\end{figure}

We consider the classical inverted pendulum system model, a widely used benchmark problem in both control and RL domain. As illustrated in Figure~\ref{fig:cartpole}, a pole is attached by a joint to a cart, which can be moving along a frictionless track. The pendulum starts upright at a random initial angle $\theta_0\in(\theta_{0,min},\theta_{0,max})$, and the goal is to prevent it from falling over by applying a force to the cart. While conceptually simple, the system dynamics are highly unstable and continuously requires fast control cycles to keep stability. When, in turn, being controlled over a wireless channel, the problem becomes an illustrative model of strict timing requirement. 
%\pp{PP: This is a bit paradoxical, since we said that latency is not suitable, but now we say that there are strict requirements. I suggest that we talk in the more general term of \emph{timing requirements} and then particularize to latency or age, when necessary.}%
 \subsection{Control System Model}
The system dynamics can be represented by the differential equations \cite{florian2007correct}:% 
%\NM{NM: citation?} \pp{PP: A single citation here could do the job for all the subsequent equations.}:%
\begin{equation}
\begin{aligned}
\ddot{\theta} = \frac{g \cdot \text{sin}(\theta) + \text{cos}(\theta) 
\left( \frac{-F-m_p l \dot{\theta}^2 \text{sin}(\theta)}{m_c + m_p} \right)}{l(\frac{4}{3} - \frac{m_p \text{cos}^2(\theta)}{m_c + m_p})}
\\
\ddot{x} = \frac{F + m_p l (\dot{\theta}^2 \text{sin}(\theta)-\ddot{\theta}\text{cos}(\theta))}{m_c + m_p} 
\end{aligned}
\label{eq:system_dynamics_cartpole}
\end{equation}%

\noindent where $x$ and $\theta$ are, respectively, the cart position coordinates and the pole angle from vertical reference. The mass of the cart is $m_c$, and the mass of the pendulum is $m_p$, while $l$ is the length of the pendulum, and $F$ is the force applied to the cart under gravity $g$. We use the Newton's notation ($\dot{\Box}, \ddot{\Box}$) to represent derivatives w.r.t time. 

%\NM{NM: What is g, $\ddot{x}$, $\ddot{\theta}$, $\dot{\theta}$?}%Note that (\ref{eq:system_dynamics_cartpole}) are nonlinear second order state variables.

By defining a state space vector $ X = [ x, \dot{x}, \theta, \dot{\theta} ]$, we can design a standard optimal 
%\NM{LQR is first time used here: define the abbrev, and explain shortly, give refs)} 
controller in two steps. First, computing the Jacobian of (\ref{eq:system_dynamics_cartpole}) around the operating point $X = [0,0,0,0]$ to linearize the plant, so that the system dynamic takes the linear time invariant form:
\begin{equation}
\begin{cases}
\dot{X} = AX + Bu + w \\
u = -KX
\end{cases}
\label{eq:system_dynamics_linear}
\end{equation}%
\noindent where $u$ is the linear state feedback control policy of gain $K$, $w$ is a process disturbance modeled as a zero-mean and one-standard deviation Gaussian white noise, $A$ and $B$ are the system transition matrix, respectively given by:%
\begin{equation}
A = \begin{bmatrix}
0 & 1 & 0 & 0 \\
0 & 0 & \frac{-12m_pg}{13m_c+m_p} & 0 \\
0 & 0 & 0 & 1 \\
0 & 0 & \frac{12(m_pg+m_cg)}{l(13m_c+m_p)} & 0 \\
\end{bmatrix}, \, 
B=\begin{bmatrix}
0  \\
\frac{13}{13m_c+m_p}  \\
0  \\
\frac{-12}{l(13m_c+m_p)}  \\
\end{bmatrix}
\label{eq:A_B}
\end{equation}%

The second step consists of finding the optimal control policy, $u^{*}$, subject to (\ref{eq:system_dynamics_linear}) that minimizes the cost function,%
\begin{equation}
J(u) = \int_{0}^{\infty}\left(X^TQX + u^TRu \right)dt,
\label{eq:lqr_cost}
\end{equation}%
%$R=0.1$ and $S= \begin{bmatrix}
%1 & 0 & 0 & 0 \\
%0 & 10 & 0 & 0 \\
%0 & 0 & 1 & 0 \\
%0 & 0 & 0 & 100
%\end{bmatrix}$
\noindent where $R$ and $Q$ are arbitrary positive defined matrices in which we can assign weights to state space variables and control signal. In control theory this kind of problem formulation is known as Linear-Quadratic-Regulator (LQR) \cite{lewis2012optimal}.

The optimal control policy then can be defined by solving the Algebraic Riccati Equation \cite{lewis2012optimal} as:%
\begin{equation}
\begin{aligned}
& A^{T}P + PA - PBR^{-1}B^{T}P+Q=0\\
& K^{*} =  R^{-1}B^{T}P \\
& u^{*} = K^{*}X \\
\end{aligned}
\label{eq:are}
\end{equation}%
For $(A,B)$ controllable, the infinite horizon LQR with ${Q,R > 0}$ gives a convergent closed-loop system \cite{lewis2012optimal}, where the stability can be easily guaranteed.%

%\section{Wireless Networked Control Model}

%\NM{NM: I proposed the title: Networked Control Model}
\subsection{Networked Control Model}
\label{ncs_model}
\begin{figure}[htpb]
    \centering
    \includegraphics[width=0.9\linewidth]{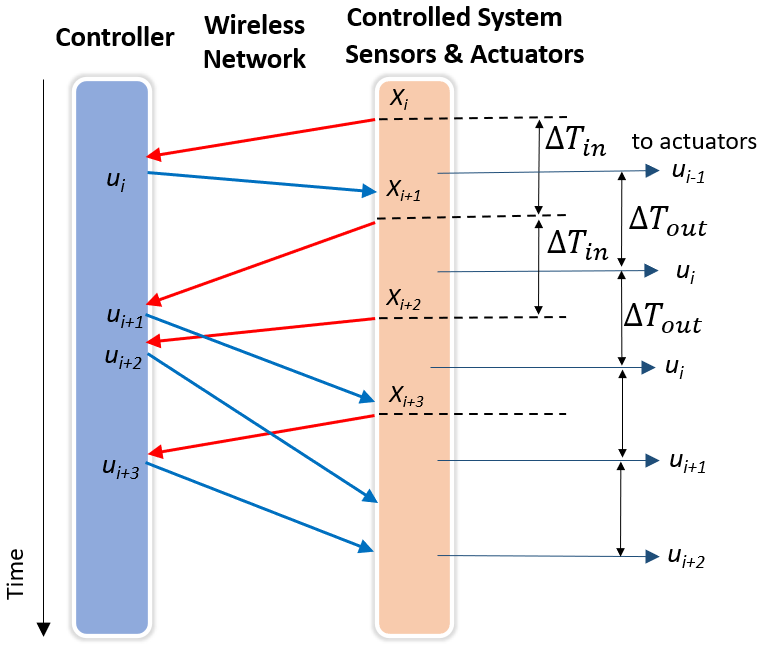}
    \caption{WNCS Model.}
    \label{fig:ncs_model}
\end{figure}

As defined in \cite{de2020wireless}, we adopt a similar NCS model to define the system behavior over the wireless medium operating in Frequency Division Duplexing (FDD) mode with separated frequency bands for the uplink (UL) and downlink (DL) directions, which makes the medium access for UL and DL independent from each other in time domain.  Figure~\ref{fig:ncs_model} illustrates the proposed model, showing the details of the interaction between the communication and application control loop.%
First, the sensor readings of the application describe the system states, $X_i$, which are stored in memory and communicated to the controller over the uplink channel. 
The readings and transmissions of sensor values are done strictly periodically with the cycle time $\Delta T_{\text{in}}$, as it is commonly done across various control systems~\cite{park2017wireless}.

At the controller, the received sensor values are also stored into the memory. The control application gets the recent values, and produces a control signal $u_i$ according to (\ref{eq:are}). Immediately after producing a control command, the controller sends it over a downlink channel to the controlled system. After finishing the current transmission, the controller  keeps waiting for the next state update from the controlled device, and starts the procedure once again. 

At the controlled system side, the received command $u_i$ is stored in the memory. The output application for actuators control (e.g., motor drives) is called periodically with the time interval $\Delta T_{\text{out}}$, calls the most recently stored command values from the memory and applies them to the application drives, producing the system dynamics of (\ref{eq:system_dynamics_cartpole}).

\subsection{Wireless channel model}
Both the DL and UL transmissions can suffer latency while delivering the information, which, in this model, depends on two main factors: the current channel quality and the total bandwidth allocated for the transmission. To evaluate this behavior, we consider the 3GPP 4-bit CQI Table 7.2.3-1~\cite{3gpp.36.213}, where we can estimate the amount of time to deliver the data considering both the channel quality indicator (CQI) and the total bandwidth allocated at the transmission. The following two assumptions have been adopted: a) the UL finishes its transmission within $\Delta T_{\text{in}}$, and b) the DL only starts a new transmission after finishing the current one. The details of the bandwidth allocation problem are discussed later in Section~\ref{prob_form}.

\subsection{System Model Discussion}
It is important to emphasize that (\ref{eq:lqr_cost}) is guaranteed to be bounded according to the Riccati-equation \cite{lewis2012optimal}. However, the combination of two main factors might affect the system LQR performance. 
%It is important to emphasize that, 
%if there is no donwlink and uplink transmission latency,
%\pp{PP: Do you mean, there is no propagation delay or...?}, 
%(\ref{eq:lqr_cost}) is guaranteed to be bounded according to the Riccati-equation %\cite{lewis2012optimal}. 
%under transmission latency, however, the system LQR performance 
%\NM{(NM: Can we be more specific, what is meant here? System stability/non-stability, risk of instability, etc?)} 
%is directly affected by the combination of two main factors. 
The first is the uplink effect, which represents the level of knowledge the controller has about the plant, meaning that, if $\Delta T_{\text{in}}$ is too high or the uplink takes overly long to deliver sensor data, 
%\pp{PP: This implies that the link has outages, but in your comm model CQI is known perfectly. How are these two things reconciled?}, 
the controller will compute the control signal using old state feedback, causing the control command to be ineffective or even harmful for the plant. The second is the downlink effect, or simply the overall delay to deliver the control signal.  This is important because if the plant applies outdated control commands for too long, the stability of the controlled system might also be compromised. 

Each of these factors might affect the plant in different ways and cannot be independently decoupled, which means that a delay in the UL will impact the DL transmission, provoking cumulative effects at the plant and at the network resources.

\section{Age of Information and Age of Loop}
\label{aol_definition}
%\bs{COMMENT: Part of this has to go to the intro}
%\PS{Moved to intro.}

%\bs{COMMENT: The confusing part of this section is that in Fig.2 everything starts with the sensor update, whereas your definition of AoL starts with the control command. Choose the same starting point and use the figure to help understand this definition. I suggest you name $g_i$ the generation times of the packet and $r_i$ the reception times of the response. }
%\PS{Figure changed.}

Age of information (AoI) provides a measure for quantifying the freshness of the knowledge we have about the status of a remote system. It represents the time duration between the generation time of the freshest received data and the current time. We can refer to its formal definition as in \cite{bedewy2019minimizing, yates2021age}, where, at time $t$, if the newest data (i.e., with the largest generation time) received at the destination was generated at time $U(t)$, the AoI~$\Delta(t)$ is defined as $\Delta(t) = t-U(t)$. 

%More formally, consider a single communication process $G = \{t_1, ..., t_i, ...\}, t_{i+1} \geq t_i$ that represents a sequence of time instances where the packets are generated at the source. Then the sequence of time instances at which those data packets are received at the destination is defined as $R = \{t_1', ..., t_i', ...\}, t_{i+1}' \geq t_i'$. We can denote $S_i = t_i' - t_i$ the time between the packet generation and reception, including all queuing and transmission delays. We can define $G(t) = \{ t_i \in G : t_i \leq T\}$ to be the set of sampling times within an arbitrary interval $[0, T]$ \NM{(shouldn't it be $t$ not $t_i$,or otherwise?)}. So, given $s[t] = \text{sup}\{t_i \in G(t) : t_i + S_i \leq t\}$ the timestamp the latest received packet was generated before the current time $t$, we define the AoI at time $t$ as $\Delta[t] = t - s[t]$.

%One of its most interesting properties is the fact that, as long as no data packet is received, the AoI linearly increases.

%However, WNCS systems like the one considered in this paper are inherently two way with a closed-loop where the UL communication can affect the DL and vice-versa, impacting either on system performance or in the use of network resources. 
The formal AoI definition, however, is inherited to a single communication link. Papers which so far explored WNCS related problems using AoI  are limited to specific analysis over only the UL \cite{champati2019performance, klugel2019aoi, gatsis2020latency} or DL \cite{huang2020wireless, liu2020latency} transmissions. However, wireless networked control systems, as the one considered in this paper, rely intrinsically on both DL and UL with a closed-loop, where the UL communication can affect the DL and vice-versa, impacting system performance and the use of network resources. A simple intuitive example that can illustrate this idea is that a high UL AoI implicates less knowledge that the controller has about the plant, which demands more urgency to deliver the control signal and, as a consequence, more network resources usage by the DL link. 
To address this implications, we propose a new metric to evaluate the overall age of a WNCS closed-loop, which we refer to as Age-of-Loop (AoL).

% extend the concept of age

%\bs{The AoI has been studied in the NCS literature. For example ... We notice that all these works define the simplest case of a one-way communication link~\cite{}, such that we have a certain AoI measure for the DL communication and another independent AoI measure for the UL communication. However, NCS systems like the one considered in this paper are inherently two way with a closed-loop where the UL communication can affect the DL and vice-versa, impacting either on system performance or in the use of network resources. The conventional one-way AoI measure is not meaningful in this case.}

Specifically, we can first verify that the state values $X_i$ are periodically generated and transmitted at time intervals of ${ t_i={\{i \cdot \Delta T_{\text{in}} \} , \forall i \in \mathbb{N}^{+}} }$, 
%for a starting time $t_0$ \NM{(can we just assume it is 0?)}
, where we can define ${ \{X_i,t_i \} }$ the sequence of generated state values and its respective time step. The control signal, in turn, is asyncrhonous and must finish the current DL transmission to start a new one upon reception of a new status update. We can define a sequence ${ \{ u_j, \hat{t}_j\} \, \forall j \in \mathbb{N}^{+} }$ with aperiodically generated control commands $u_j$ at time step $\hat{t}_j$. If ${ \{X_i,t_i \} }$ is the freshest state feedback that spawned a new control signal, we can extend the DL transmission definition to include state time information, i.e., ${ \text{DL}:\{ u_j, \hat{t}_j, t_i \} }$. Reciprocally, every state feedback also occurs under the input of the freshest control command, so that we can also extend the UL transmission definition to include control time information, i.e., ${ \text{UL}:\{ X_i, t_i, \hat{t}_j \} }$.

\begin{figure}[htpb]
    \centering
    \includegraphics[width=0.75\linewidth]{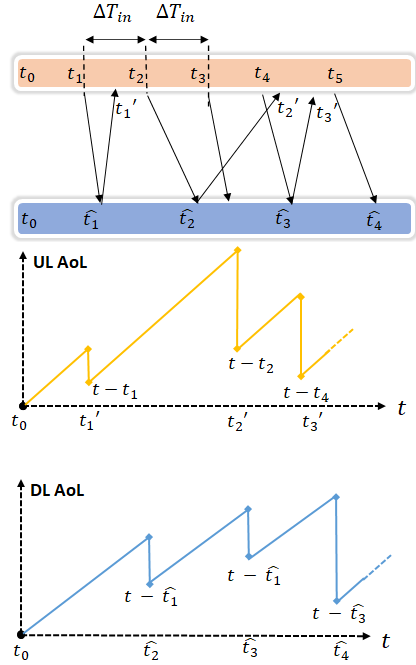}
    \caption{Time Diagram of AoL Behavior.}
    \label{fig:aoi_aol}
\end{figure}

We consider two plausible definitions of the AoL depending on the selected time origin: the DL-AoI is DL-initiated, meaning that the time origin is a new control command; the UL-AoI is UL-initiated, i.e., the time origin is a new status update in the UL. The DL AoL metric captures the time elapsed since the control command that led to the latest received update in the controller was generated. Analogously, the UL AoL metric refers to the time elapsed since the status update that caused the latest applied control command was generated at the sensor. Mathematically, if the origin is the DL, the current AoL is the difference between the current time $t$ and the time when the freshest received control command was generated:%
\begin{equation}
\text{DL AoL}(t) = t - \hat{t}_j
\end{equation}%
Likewise, if the time origin is the UL, the AoL is calculated as the difference between the current time and the time when the freshest received state was spawned:%
\begin{equation}
\text{UL AoL}(t) = t - t_i
\end{equation}%
Essentially, the main idea of AoL is to establish a metric that encompasses the behavior of two separated and locally measured entities (DL and UL) into a single instance, in which we can observe from different perspectives. 
It is important to note that, in the case of two independent AoI links, we need an instantaneous feedback to the  source  to  know  the  instantaneous  age  at  the  destination, thus making complex and potentially imprecise the union of two directions;
AoL  fixes  this. In practice, it also offers the possibility to design solutions that enclose the whole closed-loop behavior by checking the loop age from either an UL or DL perspective. For example, we can potentially design a power allocation policy for the UL by observing the current UL AoL status. Likewise, we are able to define a modulation coding scheme algorithm for the DL transmissions by observing the DL AoL. It will be proven that they are both valid to optimize the stability of the WNCS. To illustrate the proposed concept, Figure~\ref{fig:aoi_aol} shows a representative time diagram of the AoL behavior. 

\section{AoL Evaluation}
\label{aol_eval}
%\pp{PP: Explain intuitively why is there a difference when you optimize w.r.t. loop rather than the union of the two directions. One thing which is interesting about the age of the loop is that the one measuring it can measure it precisely. Note that in an AoI over a single link, we need an instantaneous feedback to the source to know the instantaneous age at the destination; AoL fixes this. } 
We aim to estimate the performance of the control system measured according to (\ref{eq:lqr_cost}) using the current AoL status calculated at the controller (DL AoL). More formally, we can use the value function definition \cite{sutton2018reinforcement} to estimate the expected LQR cost, i.e.,%
\begin{equation}
V(\Delta_{\text{AoL}}(t)) = \int_{t}^{\infty}(X^TQX+u^TRu) dt
\label{eq:V_s}
\end{equation}%
Since the control policy is unchangeable over time and the plant operation is sampled at cycles of $\Delta T_{\text{out}}$, (\ref{eq:V_s}) becomes the recursion problem:%
\begin{align}
\begin{split}
V(\Delta_{\text{AoL}}(t)) = \int\displaylimits_{t}^{t+\Delta T_{\text{out}}}
(X^TQX &+ u^TRu) dt \\ 
&+ V(\Delta_{\text{AoL}}(t+\Delta T_{\text{out}})), 
\label{eq:V_s_bellman}
\end{split}
\end{align}%
\noindent where we can solve iteratively using a temporal difference (TD) learning algorithm \cite{sutton2018reinforcement} with actual state transitions, such that:%
\begin{align}
V(\Delta_{\text{AoL}}(t)) \leftarrow V(\Delta_{\text{AoL}}(t)) + \alpha 
\Bigg[ 
\int\displaylimits_{t}^{t+\Delta T_{\text{out}}}
(X^TQX+u^TRu) dt \nonumber \\
+ \, \gamma V(\Delta_{\text{AoL}}(t+\Delta T_{\text{out}})) - V(\Delta_{\text{AoL}}(t)) \Bigg]
\label{eq:td_learning}
\end{align}%
\noindent where $\alpha$ and $\gamma$ are, respectively, the learning rate and the discount factor of future values. We can emphasize that (\ref{eq:td_learning}) converges asymptotically to the correct predictions with probability 1 if the step-size $\alpha$ decreases according to
the usual stochastic approximation conditions \cite{sutton2018reinforcement}.
\subsection{Numerical Evaluation}
Considering the following inverted pendulum configuration: ${m_c=1.0}$\,kg, ${m_p=0.1}$\,kg, ${l=0.5}$\,m, ${g=9.8}\,m/s^2$ and ${\Delta T_{\text{out}}=1}$\,ms, we evaluated the expected LQR behavior for different AoL states using (\ref{eq:td_learning}).

%Using the same methodology, we also evaluated the expected LQR cost according to the combination of DL and UL AoI, which are locally measured in each side of the communication medium.

\begin{figure}[htpb]
    \centering
    \includegraphics[width=1.0\linewidth]{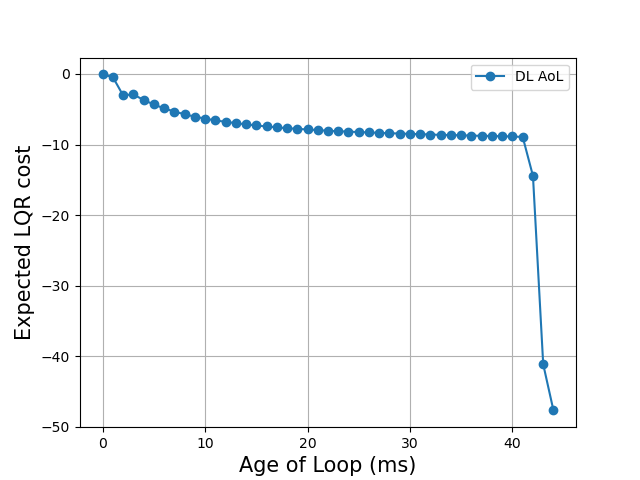}
    \caption{Expected LQR Cost vs Age of Loop.}
    \label{fig:aol_lqr}
\end{figure}

Figure \ref{fig:aol_lqr} illustrates the obtained result, where we can emphasize three extensive conclusions. First, low AoL values, as expected, provide the best system performance, such that the theoretical LQR upper bound is achieved if the AoL is close to zero. Second, prior to an AoL around $40$~ms, the LQR slightly decrease. After that point, however, the system starts to progressively become less tolerable to additional AoL delays. The third and most relevant conclusion is the fact that between $10$ and $40$~ms, there is no considerable variation at the system performance, meaning that we can avoid over-provisioning network resources by learning the system robustness.%\NM{(AoL here is an average AoL? or is AoL a constant value applied to explain and visualize its impact?)}%
\subsection{AoI vs AoL}
We performed the same numerical evaluation using a state space comprised of a single DL AoI or a single UL AoI. The goal is to verify the estimation error of the value function when we change the state space for a single AoI metric instead of AoL.
We analyze the value function estimation by verifying the Temporal Difference (TD) error, given by:%
\begin{align}
\begin{split}
\int\displaylimits_{t}^{t+\Delta T_{\text{out}}}
(X^TQX &+ u^TRu) dt \\
&+ \gamma V(\Delta_{\text{AoL}}(t+\Delta T_{\text{out}})) - V(\Delta_{\text{AoL}}(t)),
\end{split}
\end{align}
\noindent which indicates, for each state, how far the predicted value function deviates from the actual value. For example, the learning rule in (\ref{eq:td_learning}) adjusts state value in a direction that tends to reduce the TD error. 

\begin{figure}[htpb]
    \centering
    \includegraphics[width=1.0\linewidth]{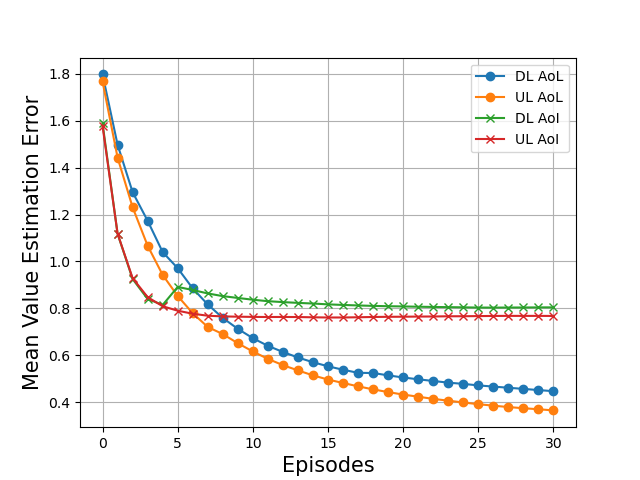}
    \caption{Mean value function estimation error over training episodes.}
    \label{fig:value_estimation}
\end{figure}

Lower TD errors indicates better accuracy about the value function estimation over each state, which is ultimately important for learning better policies, especially in RL context. Figure \ref{fig:value_estimation} illustrates the obtained result along training episodes. We can verify that, as expected, the estimated values are more precise when the whole loop age is considered. The DL and UL AoL values can be merely different, especially because of $\Delta T_{\text{in}}$.
After generation, the DL data might spend time between state transmissions before finishing the loop, 
which does not happen in UL case. Thus explaining the slight different behavior of both values in Figure~\ref{fig:value_estimation}.%
\section{Bandwidth Allocation Problem}
\label{prob_form}
As discussed in section \ref{aol_eval}, the AoL status of a WNCS can provide an estimation of the system LQR performance, so that we can use the learned value function to build a policy. In this work, we explore the bandwidth allocation problem of a remote controller, where two main objectives must be satisfied: minimize the LQR cost while using the minimum amount of bandwidth.

In more details, we can define ${B=\{b_1, b_2, \dots, b_i, \dots, b_N\}, b_{i+1}>b_i}$ a set of bandwidths in which the controller, for every DL transmission, must decide for a certain bandwidth allocation $b \in B$ given the current AoL state information and the current channel quality. So, for $T = { \{t_1, t_2, \dots, t_i, \dots t_N \}\, t_{i+1} > t_i }$ the time instances where control packets starts transmission and ${C = \{ c_1, c_2, \dots, c_i, \dots, c_N \}}$ the corresponding CQI of each transmission, the goal is to find an allocation policy ${\pi: \{\Delta_{\text{AoL}}(t_i),c_i \rightarrow b_i}\}, \, \forall t_i \in T, \, \forall c_i \in C, \, \forall b_i \in B$ that minimizes the infinite-horizon LQR cost plus the amount of bandwidth usage over the system trajectory, i.e.,%
\begin{align}
&\pi^{*} = \operatorname*{arg\,min}_{\pi} 
\left(
\int\displaylimits_{0}^{\infty}(X^TQX+u^TRu) dt +  \sum_{i=1}^{N}\frac{b_i}{b_N}
\right)  \nonumber \\
&\text{s.t.} \;
 (\ref{eq:system_dynamics_cartpole}), (\ref{eq:are})
\label{eq:policy_problem}
\end{align}%
\subsection{Solution Proposal}
\label{sol_proposal}
We can decompose the problem in (\ref{eq:policy_problem}) into sub-problems, where between two consecutive control transmissions ${[t_i, t_{i+1}), \forall t_i  \in T}$, we select at $t_i$ a bandwidth $b_i \in B$ based on the AoL and CQI state ${ \{\Delta_{\text{AoL}}(t_i), c_i\} }$. Receiving, as consequence, a one-stage decision cost of:%
\begin{equation}
\int_{t_i}^{t_{i+1}}(X^TQX+u^TRu) dt + \frac{b_i}{b_N},
\label{eq:one_stage_cost}
\end{equation}%
\noindent which depends only on the present state and the decision taken on that state. Such decision-making model is a typical
Markov Decision Proces (MDP)~\cite{sutton2018reinforcement}, where we can optimally solve each sub-problem with actual state transitions and overlap those solutions to build the overall optimal solution. In this context, we can define the following MDP configuration:
\subsubsection{State Space} Comprised of 20 AoL values, each representing regions of 5~ms from 0 to 100 ms. In addition, 15 possible CQI values for each AoL, resulting in a total of 300 states.
\subsubsection{Action Space} Represented by the bandwidth set with ten possible values: $B = { \{ 100, 200, 300, \dots, 1000\} }$~kHz.
\subsubsection{Reward} The immediate cost as defined in (\ref{eq:one_stage_cost}).
\subsubsection{Scenario} We evaluate the proposed MDP considering the NCS model described in section \ref{ncs_model}, assuming the following inverted pendulum configuration: ${m_c=1.0}$\,kg, ${m_p=0.1}$\,kg, ${l=0.5}$\,m, ${g=9.8}$\,m/s$^2$, control packet size of 1024 bits and ${\Delta T_{\text{out}}=1}$\,ms. For each run, the CQI is randomly chosen ${\{1,2,3, \dots, 15\}}$. The evaluation is also performed under different sensor feedback ${\Delta T_{\text{in}} = 1, 5, 10, 15 \text{ and } 20}$~ms. 

To solve the proposed MDP, we advocate a RL methodology for two main reasons. First, the MDP transitions probabilities are not easily tractable since the AoL variation will simultaneously depend on the channel and resource allocation of both UL and DL links. So, the UL behavior might be analytically unpredictable from the DL perspective and vice-versa. Second, learning a value function from the AoL states means that we have a prediction of system performance given the current AoL condition. In other words, this methodology offers the possibility for the network to essentially learn the control system behavior, where the bandwidth allocation policy is just one of multiple network functions in which it can serve. We could easily extend the learned values to find optimal polices, for example, in terms of packet length, power allocation, antenna configuration and so on. 

Hence, we solved the proposed MDP by applying a TD RL algorithm, based on a $\epsilon$-greedy decision making during training, with exponential learning and exploration rate decay. \cite{sutton2018reinforcement}. 

%\pp{PP: You need to put more justification why Reinforcement Learning is needed, i.e. why this problem cannot be addressed with a direct analysis.} 

\section{Results}
\label{res}
We compare the proposed solution with a bandwidth allocation scheme based on predefined delay requirements, which is the general solution currently used in industry. In more details, given an arbitrary requirement of $T_r$~ms for the control packet to be delivered, we can directly calculate the minimum amount of bandwidth to achieve the necessary requirement using the 3GPP 4-bit CQI Table 7.2.3-1~\cite{3gpp.36.213} and the total packet size. These baseline approaches, as well as the RL scheme, were evaluated on the scenario described in Section~\ref{prob_form}.

\begin{figure}[htpb]
    \centering
    \includegraphics[width=1.0\linewidth]{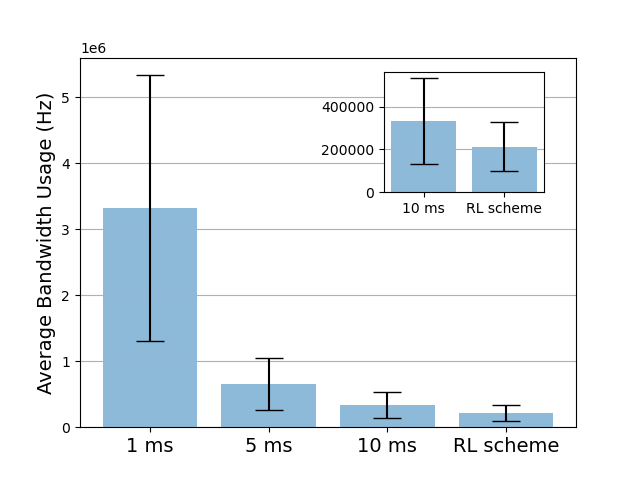}
    \caption{Total amount of bandwidth usage for each method.}
    \label{fig:bw_usage}
\end{figure}

\begin{figure}[htpb]
    \centering
    \includegraphics[width=1.0\linewidth]{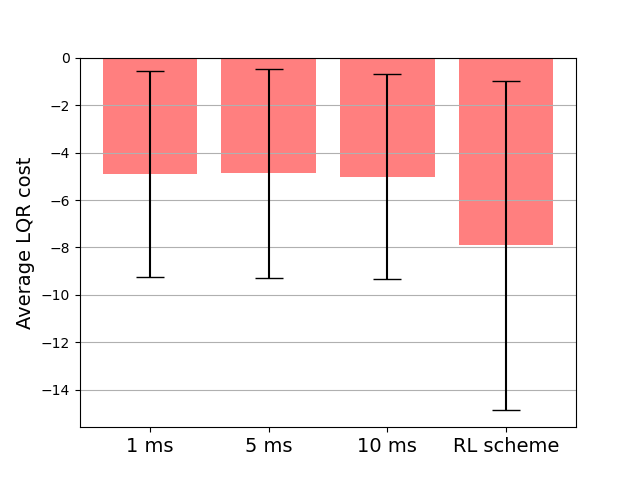}
    \caption{Total amount of LQR cost for each method.}
    \label{fig:lqr_cost}
\end{figure}

We analyze the results for three common network requirements, ${T_r=1 \, \text{ms}}$, ${T_r=5 \, \text{ms}}$ and ${T_r=10}$~ms. In each case, we analyzed the total bandwidth usage and the total LQR cost, which are respectively illustrated in Figure \ref{fig:bw_usage} and Figure \ref{fig:lqr_cost}.%

The immediate conclusion we can verify is that the RL scheme was capable to learn the system delay requirement, such that we can relate the LQR cost in Figure \ref{fig:lqr_cost} with the result in Figure \ref{fig:aol_lqr} to show that it is operating around the LQR edge performance (around -8) in order to save bandwidth. The second conclusion is that, as expected, strict latency requirement (${T_r=1}$~ms) demands more bandwidth usage. Compared to ${T_r=10}$~ms, however, the RL scheme could still save $36$\% more bandwidth, which is an indication that $10$~ms is still a sub-optimal requirement, but we can learn it from the RL algorithm. 

%\NM{NM:(Does a lower LQR value mean that the system is in general more stable or unstable? I think more comments to this is still need for better understanding).}
%\PS{Flexibility to prioritize LQR instead?}

\section{Conclusions}
In this work, we proposed a new metric to evaluate the age of an WNCS closed-loop, and we applied this metric, the Age of Loop, to track the LQR performance of a inverted pendulum control system. Furthermore, we also propose a bandwidth allocation policy based on the age of loop and channel quality information, showing that we can learn the system robustness in order to avoid over-provisioning of network resources on a networked control system.

As future works, we intend to explore a joint DL and UL RL methodology where both cooperate to optimize system performance and network resources.

%\section*{Acknowledgement}
%This project has received funding from the European Union’s Horizon 2020 research and innovation program under the Marie Skłodowska-Curie grant agreement No 813999. Further information is available at https://windmill-itn.eu/
%This document reflects the views of the author(s) and does not necessarily reflect the views or policy of the European Commission. The REA cannot be held responsible for any use that may be made of the information this document contains.

\bibliography{refs}
\bibliographystyle{IEEEtran}

\end{document}